\documentclass[useAMS,usenatbib,longnamesfirst,usegraphicx]{mn2e}
\usepackage{txfonts}
\usepackage{color}

\newcommand{\etal}{et al.}
\newcommand{\Chandra}{{\it Chandra}}
\newcommand{\casa}{\textrm{CANS}}

\newcommand{\NH}{N_{\mathrm{H}}}

\begin{document}
\title[Young Cas~A Neutron Star]{Cooling rates of neutron stars and
the young neutron star in the Cassiopeia~A
supernova remnant}

\author[D. G. Yakovlev \etal]{Dmitry G.
 Yakovlev$^{1}$\thanks{E-mail: yak@astro.ioffe.ru},
 Wynn C. G. Ho$^{2}$\thanks{E-mail: wynnho@slac.stanford.edu},
 Peter S. Shternin$^{1}$,
 Craig O. Heinke$^{3}$, 
\newauthor
 and Alexander Y. Potekhin$^{1,4}$ \\ 
$^{1}$Ioffe Physical Technical Institute,
 Politekhnicheskaya 26, 194021 St.\ Petersburg, Russia \\
$^{2}$School of Mathematics, University of Southampton,
 Southampton, SO17 1BJ, United Kingdom \\
$^{3}$Department of Physics, University of Alberta,
 Room 238 CEB, 11322-89 Avenue, Edmonton, AB, T6G 2G7, Canada \\
$^{4}$Isaac Newton Institute of Chile, St.~Petersburg Branch,
St.~Petersburg, Russia}

\date{Accepted . Received ; in original form}
\pagerange{\pageref{firstpage}--\pageref{lastpage}} \pubyear{2010}
\maketitle
\label{firstpage}

\begin{abstract}
We explore the thermal state of the neutron star in the
Cassiopeia~A supernova remnant using the recent result of
\citet{hoheinke09} that the thermal radiation of this star is
well-described by a carbon atmosphere model and the emission comes
from the entire stellar surface. Starting from neutron star
cooling theory, we formulate a robust method to extract neutrino
cooling rates of thermally relaxed stars at the neutrino cooling
stage from observations of thermal surface radiation.  We show how
to compare these rates with the rates of standard candles -- stars
with non-superfluid nucleon cores cooling slowly via the modified
Urca process.
We find that the internal temperature of standard candles is a
well-defined function of  the stellar compactness parameter
$x=r_g/R$,
irrespective of the equation of state of neutron star matter
($R$ and $r_g$ are circumferential and gravitational radii, respectively).
We demonstrate that the data on the Cassiopeia~A
neutron star
can be explained in terms of three
parameters: $f_\ell$, the neutrino cooling efficiency with respect
to the
standard candle;
the compactness $x$; and the amount of light elements in the heat
blanketing envelope. For an ordinary (iron) heat blanketing
envelope or a low-mass ($\lesssim 10^{-13}\,M_\odot$) carbon
envelope, we find the efficiency $f_\ell \sim 1$ (standard
cooling) for $x \lesssim 0.5$ and $f_\ell \sim 0.02$ (slower
cooling) for a maximum compactness $x\approx 0.7$.  A heat blanket
containing the maximum mass ($\sim 10^{-8}\,M_\odot$) of light
elements increases $f_\ell$ by a factor of 50. We also examine the
(unlikely) possibility that the star is still thermally
non-relaxed.
\end{abstract}


\section{Introduction} \label{sec:intro}

Neutron stars are born with very high internal temperatures ($T
\gtrsim 10^{11}$~K) but gradually cool down.
Observing the thermal radiation from cooling isolated neutron stars
and comparing their temperatures to theoretical models allows one
to explore the (still uncertain) physical properties of superdense matter
in neutron star cores
\citep[see, e.g.,][for review]{pethick92,tsuruta98,yakovlevpethick04,pageetal06,pageetal09}.

The supernova remnant Cassiopeia~A (Cas~A),
at a distance of $d=3.4^{+0.3}_{-0.1}$~kpc \citep{reedetal95}, is one
of the youngest-known in the Milky Way. The supernova that gave rise
to the remnant may have been observed in 1680 \citep{ashworth80},
though this may have been a mis-identification \citep[see,
e.g.,][for review and
discussion]{thorstensenetal01,stephensongreen02}; nevertheless, a
supernova explosion in $1681\pm 19$ years has been estimated by examining
the expansion of the remnant \citep{fesenetal06}. Hereafter we shall
assume that Cas~A has a current age of $330\pm 20$~yr. Although the
supernova remnant is extremely well-studied, the central compact
source was only identified in first-light \Chandra\ X-ray
observations \citep{tananbaum99} and subsequently studied by
\citet{pavlovetal00,chakrabartyetal01,pavlovluna09}. Recently, it
was shown \citep{hoheinke09} that the compact source in the remnant
is a neutron star, which we will refer to as \casa\ (for Cas~A
Neutron Star or CArbon Neutron Star), with a carbon atmosphere and
low magnetic field ($B\lesssim 10^{11}$~G). From the spectral
fitting of the \Chandra\ observations, the neutron star
(gravitational) mass $M$, (circumferential) radius $R$, and
effective (non-redshifted) surface temperature $T_\mathrm{s}$ were
found to be $M\approx 1.5-2.4M_\odot$, $R\approx 8-18$~km (at
$90\%$~confidence), and $T_\mathrm{s} \sim 2\times 10^6$~K,
respectively \citep{hoheinke09}. Prior to this last work, the
possibility to study the \casa\ thermal evolution was hindered by
the assumption that the measured temperature reflected a
local hot spot, hence it could only be used to set an upper limit on
the surface temperature \citep{pavlovetal00}. However, with the
recent results showing that the X-ray emission arises from the
entire neutron star surface, $T_\mathrm{s}$ can be treated as the
average surface temperature. Because the next youngest neutron
stars, for which surface thermal emission has been detected, have
ages exceeding a thousand years, the \casa\ serves as a valuable
window into the early life of a cooling neutron star.

In this paper we use the theory of neutron star cooling and
analyze the thermal state of the \casa. Because cooling theory
gives a wealth of cooling scenarios involving many uncertain
properties of superdense matter in neutron star interiors (such as
composition, equation of state -- EOS, superfluid properties of
baryons, etc.), we do not test these scenarios one by one.
Instead, we propose a robust method to analyze observations of
neutron stars of age $\sim 10^2 - 10^5$~yr, i.e., those which have
thermally relaxed interiors and cool via neutrino emission from
their cores. We describe the procedure to extract the neutrino
cooling rate of neutron stars (the ratio of their neutrino
luminosity to the heat capacity) from observations and to compare
the measurements with the theoretical cooling rates of neutron
stars with non-superfluid nucleon cores which cool via the
modified Urca process of neutrino emission. These theoretical
cooling rates can be treated as standard neutrino candles, and the
comparison between observations and theory can give valuable
information on the physical properties of neutron star interiors.
We apply this method to interpret observations of the \casa,
evaluate the \casa\ cooling rate and outline various physical
models of \casa\ internal structure that are compatible with the
observations. We also explore the possibility that the \casa\ is
still in a thermally non-relaxed state.

\section{Data Analysis} \label{sec:obs}

We summarize here the observations and spectral fitting of the
\casa\ used in our study; further details can be found in
\citet{hoheinke09} and \citet{heinkeho10}. Sixteen sets of \Chandra\ X-ray
Observatory archival data are considered (see \citealt{heinkeho10} for a
listing), all using the ACIS-S charge-coupled device which provides spatial
and spectral information \citep{garmireetal03}. A series of observations,
totalling 1~megasecond (the majority of all extant data), was taken in
2004 to study the supernova remnant \citep{hwangetal04}. Several
medium-length ($\sim$50 ks) observations have been taken at various
times between 2000 to 2009.  A 70-kilosecond observation
in 2006 was designed to study the compact source \citep{pavlovluna09},
using a subarray
configuration to reduce instrument pileup \citep{Davis01}.  As discussed in
\citet{heinkeho10}, most of the 2004 dataset is subtly
contaminated by the presence of bad pixels affecting the flux from \casa.
Although Chandra response functions include the effect of the
bad pixels, the calibration becomes
rather different, inducing small ($<2$\%) variations in the computed
temperature, depending on telescope orientation (as the position of the
\casa\ with respect to the bad pixels varies).  We choose to include
all data and permit variations in the fitted temperatures, while
forcing all other parameters
to be the same between different observations (except the pileup
grade-migration parameter $\alpha$, permitted to vary with different
frame times).
We justify including all data by the much stronger constraints we obtain on
neutron star mass and radius when we include all the data, and by
the fact that the temperature variations between observations are
small (thus any effects on, e.g., the inferred neutron star
mass and radius, will be $\ll 10$\%).
CIAO~4.1 and XSPEC~12.4.0 are used for the data reduction and analysis.

We note that using seven \Chandra\ ACIS observations from the past
ten years,
\citet{heinkeho10} showed that the temperature of the \casa\ has
been declining.  However, the relative change is only $\approx 4\%$
during the period of observation.  Since we are considering the
\casa\ in the context of its
long-term temperature evolution in
the present work, we treat the observed temperature as a constant at
the current epoch.

To fit the spectrum of the \casa, we used neutron star atmosphere
models with various compositions \citep{hoheinke09}. These models
are constructed assuming a plane-parallel atmosphere (since the
atmosphere thickness  $\sim 10$~cm is much smaller than the stellar
radius). The atmosphere is in hydrostatic and radiative equilibrium
at constant surface gravitational acceleration
$g=GM/(R^2\sqrt{1-x})$, where
\begin{equation}
   x={r_g \over R}={2GM \over c^2R}=
   0.2953\, {M \over M_\odot}\,{10~{\rm km} \over R}
\label{eq:x}
\end{equation}
is the stellar compactness parameter, i.e., the ratio of its Schwarzschild
radius, $r_g=2GM/c^2$, to $R$. The efficient separation of light and
heavy elements results in atmospheres composed of a single element
\citep{alcockillarionov80,brownetal02}; the opacities are obtained
from tables computed by the Opacity
Project\footnote{http://cdsweb.u-strasbg.fr/topbase/op.html}.
Further details of the atmosphere model construction are given in
\citet{holai01,hoheinke09}. Only a
non-magnetic carbon atmosphere provides both a good fit to the spectrum
of the \casa\ and an emission size consistent with theoretical predictions
for neutron star radii \citep{hoheinke09}.
In addition,
 \citet{changbildsten04} and \citet{changetal10}
  find that the high
temperatures present in young ($\lesssim 10^3$~y) neutron stars
remove all surface hydrogen and helium on timescales shorter
than the age of the neutron star.
The (minimum) column depth and total mass of carbon required to form an
optically thick photosphere is $\approx 50\mbox{ g cm$^{-2}$}$ and
$\sim 10^{15}$~g, respectively, and the density at one optical depth is
$\sim 0.02-4\mbox{ g cm$^{-3}$}$.
Thus we only consider hereafter the carbon atmosphere model.

Fits of the \casa\ spectrum by a neutron star atmosphere model
depend on several parameters (in addition to corrections from
pile-up and dust scattering). They are neutron star radius $R$ and
mass $M$, effective surface temperature $T_\mathrm{s}$, and
effective column density of interstellar hydrogen $\NH$. In our
spectral fits we have fixed the distance to $d=3.4$~kpc
\citep{reedetal95}. The smallest uncertainty is in $\NH$ (to within
$\sim 15\%$ around $\NH=1.8\times 10^{22}$~cm$^{-2}$).
The measured flux implies a
bolometric (1~eV$-$10~keV) luminosity of $7.5^{+3.2}_{-1.2}\times
10^{33}$~erg~s$^{-1}$ at 90\% confidence, assuming a distance
between 3.3$-$3.7~kpc.

\begin{figure}
\includegraphics[width=\columnwidth,bb = 20 150 555 685]{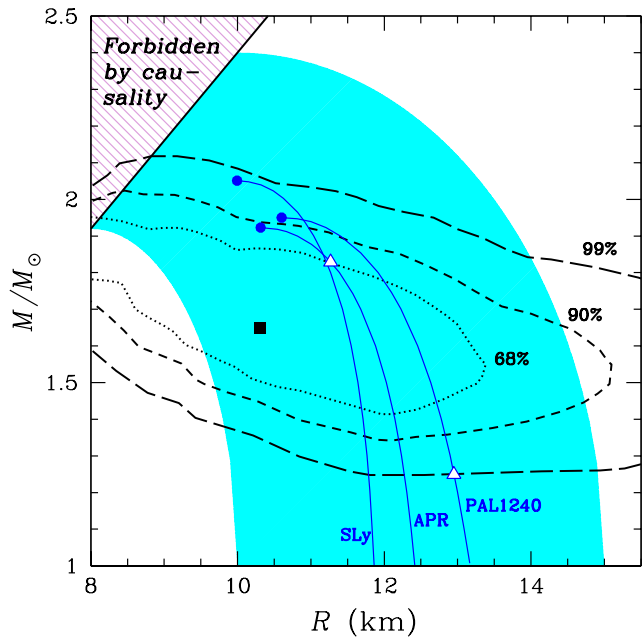}
\caption{(Color on line) \casa\ mass versus radius as inferred
from spectral fits at 68\% (dotted line), 90\% (short-dashed line)
and 99\% (long-dashed line) confidence. The filled square gives
the best-fit values. The shaded region in the upper left corner is
forbidden by the causality principle \citep[e.g.,][]{hpy07}. The
lower shaded region approximately shows the allowed values of $M$
and $R$ for realistic neutron star models. The three solid lines
show $M-R$ relations for neutron stars with the SLy, APR and
PAL1240 EOSs. Filled dots indicate the most massive stable neutron
stars for these
EOSs. Triangles are the stellar masses above which the direct Urca process
becomes allowed for the APR and PAL1240 EOSs (this process
is impossible in a stable star for the SLy EOS).
 \label{fig:mr} }
\end{figure}

\begin{figure*}
 \includegraphics[width=\textwidth]{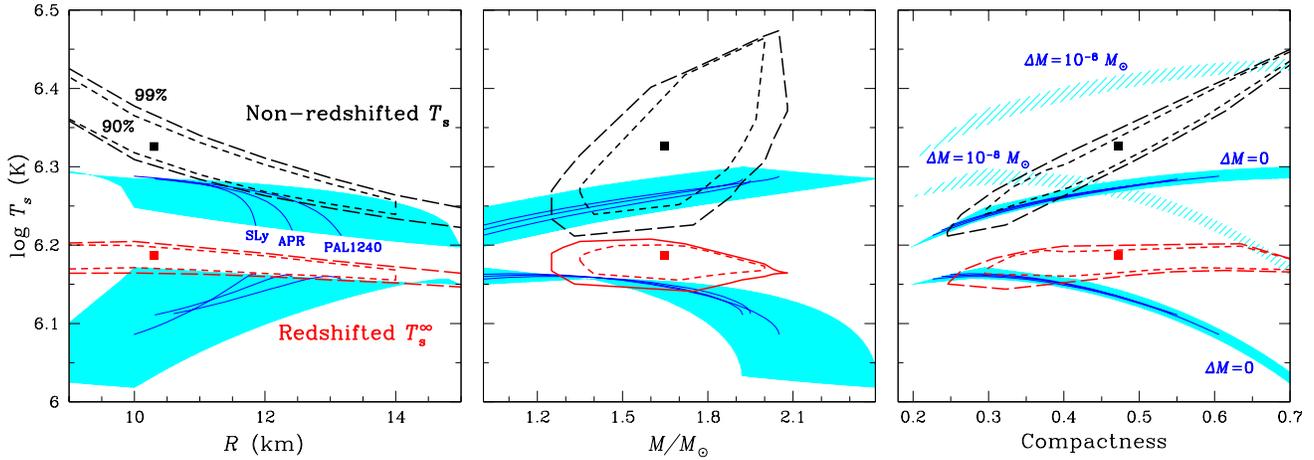}
\caption{(Color on line) Non-redshifted and redshifted effective
surface temperatures $T_\mathrm{s}$ and $T_\mathrm{s}^\infty$ (upper
and lower lines, respectively) of the \casa, inferred from
observations at 90\% (short-dashed lines) and 99\% (long-dashed lines)
confidence levels, as a function of radius (left), mass (middle) and
compactness (right). The shaded regions are theoretical values of
$T_\mathrm{s}$ and $T_\mathrm{s}^\infty$
for neutron star standard candles
(having non-superfluid nucleon cores and cooling via the modified
Urca process) with realistic $M$ and $R$ from the shaded region in
Fig.~\ref{fig:mr}. The solid lines refer to
neutron star models constructed using the SLy, APR and PAL1240
EOSs
with the direct Urca process switched off.  All theoretical
surface temperatures are calculated assuming iron heat blanketing
envelopes, except for the two lightly-shaded regions in the right
panel which assume carbon envelopes of mass $\Delta M =
10^{-8}M_\odot$. \label{fig:ts} }
\end{figure*}

Spectral fits give us allowed regions in $T_\mathrm{s}$, $M$ and
$R$ parameter space at
different confidence levels. These results
are shown in Figs.\ \ref{fig:mr} and \ref{fig:ts}. The best-fit
values are $\log T_\mathrm{s}~\mathrm{[K]}=6.326$, $M=1.65\,
M_\odot$ and $R=10.3$ km (shown by filled squares). Fig.\
\ref{fig:mr} presents 68\% (dotted lines), 90\% (short-dashed
lines) and 99\% (long-dashed lines) contours on the mass-radius
diagram for neutron stars. The shaded upper-left corner is
forbidden by the causality principle. The $M-R$ diagram and
causality constraints are discussed, e.g., in
\citet*{hpy07}. The $M-R$ uncertainties are seen to be wide.

To compare with theory, the solid lines
in Fig.~\ref{fig:mr} give theoretical $M(R)$ relations for neutron stars
which contain nucleon cores
with
three
different
EOSs:
SLy EOS by \citet{douchinhaensel01};
APR EOS by \citet{akmaletal98};
a version PAL1240 of the PAL EOS by \citet{prakashetal88}.
By APR we mean the parametrization of APR results suggested by
\citet{hhj99}; specifically, we use version APR~I proposed by
\citet{gusakovetal05}. The PAL1240 is a version of the PAL EOS with
the compression modulus of saturated nuclear matter $K_0=240$~MeV
and model 1 for symmetry energy \citep{prakashetal88}.
Filled dots on the solid lines indicate maximum-mass neutron
star configurations ($M=M_\mathrm{max}$). Triangles correspond to
the mass thresholds $M_\mathrm{DU}$ where the powerful direct Urca
process of neutrino emission becomes allowed (more massive stars
cool rapidly if the direct Urca process in their cores is not suppressed by
superfluidity; see Sec.~\ref{sec:neutrinostage}).
For the SLy EOS we have $M_\mathrm{max}=2.05\, M_\odot$, and rapid
cooling is forbidden
at any $M\leq M_\mathrm{max}$. For the APR EOS,
$M_\mathrm{DU}=1.829\, M_\odot$ and
$M_\mathrm{max}=1.929\,M_\odot$, while for the PAL1240 EOS,
$M_\mathrm{DU}=1.25 M_\odot$ and $M_\mathrm{max}=1.95 M_\odot$.
The central shaded region in Fig.~\ref{fig:mr} shows the range of
$M$ and $R$ that can be obtained for other EOSs which we consider
as realistic. We see that the  \casa\ limits on $M$ and $R$,
inferred from observations, are wider than the
range of theoretical predictions.

In Fig.~\ref{fig:ts} we present contours of $T_\mathrm{s}$ inferred
from the \casa\ observations as a function of $R$ (left), $M$
(middle) and $x$ (right). These contours are given by the upper
curves. They are projections of the appropriate $T_\mathrm{s}-M-R$
regions (at 90\% and 99\% confidence levels)\footnote{Note that
contours in Fig.~\ref{fig:mr} refer to two-dimensional ($M-R$)
confidence levels, while contours in
Figs.~\ref{fig:ts}--\ref{fig:fl} are for somewhat different
three-dimensional ($T_\mathrm{s}-M-R$) confidence levels. However,
this difference does not affect our principal conclusions.} on the
corresponding axis, $R$, $M$ or $x$. Having $T_\mathrm{s}-M-R$
regions, we also show the associated $T_\mathrm{s}^\infty-M-R$ regions,
where $T_\mathrm{s}^\infty=T_\mathrm{s} \, \sqrt{1-x}$ is the
(redshifted) surface temperature as measured by distant observer. The lower
curves in Fig.~\ref{fig:ts} are the corresponding contours of
$T_\mathrm{s}^\infty$. Other curves are explained and discussed in
Section~\ref{sec:candle}.

\section{Three cooling stages}
\label{sec:cooling}

Generally, neutron stars cool via neutrino emission from their
interiors (mainly from the core) and via heat conduction to the
surface and successive thermal surface emission.  Cooling theory
solves the heat conduction problem within the star in General
Relativity \citep{thorne77}, accounting for internal neutrino
energy losses and surface emission of photons. One can distinguish
three main cooling stages.  During the first stage (which lasts
from $\sim 10$ yr to a few centuries depending on the specific
model), a newly born neutron star is thermally non-relaxed, with
the core being cooler than the crust because of stronger neutrino
emission in the core \citep[see][for
details]{nomototsuruta81,nomototsuruta87,
lattimeretal94,gnedinetal01,shterninyakovlev08}. The core is
thermally decoupled from the crust; the surface temperature
$T_\mathrm{s}$ reflects the physics of the crust. During the
second cooling stage (which lasts for $10^5-10^6$ yr), the star is
thermally relaxed inside; it cools mainly via neutrino emission.
The main temperature gradient is located in a thin layer near the
surface (in the so called heat-blanketing envelope, at densities
$\rho < \rho_\mathrm{b} \approx 10^{10}$ g~cm$^{-3}$,
\citealt{gudmundssonetal83}).  During the third (final) stage, the
thermally relaxed star cools via the surface emission of thermal
photons. During each cooling stage, the thermal radiation of
neutron stars carries different information on the
stellar structure.
Isolated neutron stars,
whose thermal surface radiation has been detected, are mostly
at
the (second) neutrino cooling stage (with isothermal interior).
The \casa\ is the youngest of these neutron stars and is thus of
special interest. We study the thermal state of this neutron star
using our general relativistic cooling code \citep{gnedinetal01}
and useful semi-analytic expressions described below.

\section{Neutrino cooling of thermally relaxed stars}
\label{sec:stage2}

\subsection{Basic relations}
\label{sec:relaxed}

A thermally relaxed star has an isothermal interior which extends from
the center to the heat blanketing envelope
(i.e., at $\rho\gtrsim\rho_\mathrm{b}$).
Taking into account the effects of General
Relativity (e.g., \citealt{thorne77}), isothermality means
spatially constant redshifted internal temperature
$\widetilde{T}(t)=T(r,t) \exp(\Phi(r))$, where $T(r,t)$ is the local
internal temperature, $r$ is radial coordinate and $\Phi(r)$ is the
metric function that determines gravitational redshift. The cooling
equations in such a star are much simpler than the general equations
of heat transport. They reduce to the equation of global thermal
balance \citep{gs80},
\begin{equation}
   C(\widetilde{T}) \, { {\rm d} \widetilde{T} \over {\rm d} t}=
   -L_\nu^\infty(\widetilde{T}) - L_\mathrm{s}^\infty(T_\mathrm{s}),
\label{eq:therm}
\end{equation}
where $L_\nu^\infty(\widetilde{T})$ and
$L_\mathrm{s}^\infty(T_\mathrm{s})$ are, respectively, the neutrino
luminosity and the photon thermal luminosity of the star (redshifted
to a distant observer), while  $C(\widetilde{T})$ is the stellar heat
capacity. These quantities are given by
\begin{eqnarray}
   L_\nu^\infty(\widetilde{T})& = & \int {\rm d}V \, Q_\nu(T,\rho) \exp(2 \Phi(r)),
\label{eq:Lnu} \\
   L_\mathrm{s}^\infty(T_\mathrm{s}) & = & 4 \pi \sigma  T_\mathrm{s}^4 R^2\,
   \left(1- x \right),
\label{eq:Ls}   \\
   C(\widetilde{T})& = & \int {\rm d}V \, C_V(T,\rho),
\label{eq:C}
\end{eqnarray}
where $Q_\nu(T,\rho)$ is the neutrino emissivity,
$C_V(T,\rho)$ is the specific heat capacity, $\sigma$ is the
Stefan-Boltzmann constant, and ${\rm d}V=4 \pi r^2 \,{\rm d}r\,
\exp(\lambda)$ is the element of proper volume determined by
the appropriate metric function $\lambda(r)$.

The hydrostatic stellar structure [$\rho(r)$, $\Phi(r)$, etc.] is
calculated by solving the Tolman-Oppenheimer-Volkoff equation (e.g.,
\citealt{st83}). Then the cooling problem can be easily solved by
integrating (\ref{eq:therm}); the solution gives $\widetilde{T}(t)$
and $T_\mathrm{s}(t)$. To this aim, one needs to calculate the two
functions, $L_\nu^\infty(\widetilde{T})$ and $C(\widetilde{T})$ [by
integrating over the stellar volume in (\ref{eq:Lnu}) and
(\ref{eq:C})]. In addition, one needs to relate $T_\mathrm{s}$ to
the temperature $T_\mathrm{b}$ at the bottom of the heat blanketing
envelope ($\rho=\rho_\mathrm{b}$) by solving separately the
stationary one-dimensional heat conduction problem within this
envelope (e.g.,
\citealt{gudmundssonetal83,potekhinetal97,potekhinetal03}).
Approximately, $T_\mathrm{s} \propto T_\mathrm{b}^{1/2}$
\citep{gudmundssonetal83}. Since the envelope is thin, one can set
$\exp(\Phi(r)) \approx \exp(\Phi(R))=\sqrt{1-x}$ within it, so that
$\widetilde{T} \approx T_\mathrm{b} \sqrt{1-x}$.  Eq.\
(\ref{eq:therm}) is solved with some initial condition,
$\widetilde{T}(0)=\widetilde{T}_{0}$. However, for realistic values
$\widetilde{T}_{0} \gtrsim 10^9$~K, the dependence
$\widetilde{T}(t)$ at $t \gtrsim 1$ yr is insensitive to
$\widetilde{T}_{0}$ (the initial condition is forgotten at such
$t$).

Thus the cooling of thermally relaxed neutron stars is governed
by the functions $L_\nu^\infty(\widetilde{T})$, $C(\widetilde{T})$
and $T_\mathrm{s}(T_\mathrm{b})$,
which are the only three functions that can be tested by comparing
neutron star cooling theory with observations.
This cooling theory is fairly insensitive to the microphysics of the
stellar crust (at $\rho \gtrsim \rho_\mathrm{b}$) and to the thermal
conductivity in the isothermal interior.
The functions $L_\nu^\infty(\widetilde{T})$ and $C(\widetilde{T})$ are
sensitive to the (uncertain) microphysics in the neutron star core,
while the $T_\mathrm{s}(T_\mathrm{b})$-relation is based on the
(better-known) plasma physics in the outer neutron star envelope.

Our cooling model implies that the isothermal stellar interior
($\rho \geq \rho_\mathrm{b}$) is spherically symmetric. The
structure of the heat blanketing envelope  ($\rho <
\rho_\mathrm{b}$)  can deviate from spherical symmetry under the
effects of strong magnetic fields. If the latter effects are
substantial, $T_\mathrm{s}$ varies over the neutron star surface,
which is taken into account in our envelope models
 \citep{potekhinetal03,potekhinetal07}.
However, we have checked that the magnetic field $B \lesssim
10^{11}$~G, which can be present in the surface layers of the \casa,
does not affect
$L_\mathrm{s}^\infty$, averaged over the surface.
  Thus, we set $B=0$, so
that the envelope models are determined only by the surface gravity
$g$ and by the mass  $\Delta M$ of light (accreted) elements. This
assumption is supported by the non-detection thus far of pulsations from the
\casa\
\citep{murray02,mereghetti02,ransom02,pavlovluna09,halperngotthelf10}.

Our basic model will be the standard iron heat blanketing envelope. The
presence of a small amount of carbon ($\Delta M \sim
10^{-18}M_\odot$) in the \casa\ atmosphere does not affect the
cooling properties of the star. We also consider models with more substantial
light-element envelopes. Light elements increase the thermal
conductivity and make the envelope more heat transparent (increase
$T_\mathrm{s}$ for the same $T_\mathrm{b}$).
Since we assume a
carbon atmosphere, the envelope can contain carbon or heavier
elements (hydrogen or helium, if present, would flow up to the surface). To this
aim, we have constructed new (carbon-iron) models with different
carbon mass $\Delta M$, and iron under the carbon layer. The
thickest carbon layer, with $\Delta M \sim 10^{-8}M_\odot$, can
extend to the density of a few times of $10^9$ g~cm$^{-3}$; at
higher densities carbon is destroyed by beta captures and
pycnonuclear reactions.

\subsection{Neutrino cooling rate}
\label{sec:neutrinostage}

The cooling problem is further simplified during the neutrino cooling
stage. In this case $L_\mathrm{s}^\infty$ can be neglected in
(\ref{eq:therm}), so that
\begin{equation}
    {{\rm d} \widetilde{T} \over {\rm d} t}= - \ell(\widetilde{T}),
    \quad \ell(\widetilde{T}) \equiv { L_\nu^\infty(\widetilde{T})
    \over C(\widetilde{T}) }.
\label{eq:cool2}
\end{equation}
The internal temperature is controlled by the physics of the
stellar interior (mainly, the core), being insensitive to the
structure of the heat blanketing envelope.
The cooling is governed by the one function
$\ell(\widetilde{T})$, which specifies the cooling rate -- the ratio of
the neutrino luminosity to the heat capacity [K~yr$^{-1}$] -- that
is mainly determined by the properties of superdense core. It is
only $\ell(\widetilde{T})$ that can (in principle) be extracted from
measured values of $T_\mathrm{s}(t)$.
However, the extraction is not
straightforward.
Two stars with the same
mass and internal structure have the same $\widetilde{T}(t)$ (look
the same from inside). However, if their heat blanketing envelopes
are different, they have different surface temperatures
$T_\mathrm{s}(t)$ (look different from outside).

We discuss the extraction, neglecting the effects of very fast
rotation, superstrong internal magnetic field and possible
deviations from beta-equilibrium. Then there is a single unique cooling
function $\ell(\widetilde{T})$ for the real EOS and
microphysics
of neutron star cores. This unknown function depends only on
$\widetilde{T}$ and $M$ (or some other quantity, $R$ or $x$,
instead of $M$). Theoretical models give different functions
$\ell(\widetilde{T})$, for different compositions, EOSs, and
superfluid properties of the core. Of the two ingredients of
$\ell(\widetilde{T})$ [$L_\nu^\infty(\widetilde{T})$ and
$C(\widetilde{T})$], the neutrino luminosity
$L_\nu^\infty(\widetilde{T})$ is much more sensitive to the
microphysics of superdense matter, than the heat capacity
$C(\widetilde{T})$. Therefore, $\ell(\widetilde{T})$ mostly
reflects the neutrino luminosity of neutron star cores. The
standard neutrino emission mechanism is the modified Urca process
(e.g., \citealt{pethick92,yakovlevetal01,pageetal06,pageetal09})
in non-superfluid cores. It gives $L_{\nu \rm
MU}^\infty(\widetilde{T}) \sim (10^{30} - 10^{31}) \,
\widetilde{T}_8^8$ erg~s$^{-1}$ (where
$\widetilde{T}_8=\widetilde{T}/10^8~{\mathrm K}$). We consider
this as a neutrino standard candle (standard slow neutrino
emission). The heat capacity of such cores is mainly determined by
neutrons, $C(\widetilde{T}) \sim 10^{38}\,\widetilde{T}_8$
erg~K$^{-1}$. The cooling rate of the standard candle (SC) can be
estimated as
\begin{equation}
     \ell_{\rm SC}(\widetilde{T}) \sim (0.3-3)\,
     \widetilde{T}_8^7~~\mathrm{K~yr}^{-1}.
\label{eq:ell-sc}
\end{equation}
We discuss $\ell_{\rm SC}(\widetilde{T})$ in more detail in Section
\ref{sec:candle}.

The actual (unknown) cooling rate can strongly differ from the
standard one, either enhanced or reduced. In this respect it is
useful to introduce the ratio
\begin{equation}
     f_\ell
     = \ell(\widetilde{T}) / \ell_{\rm SC}(\widetilde{T}),
\label{eq:f-ell}
\end{equation}
which expresses the actual rate $\ell(\widetilde{T})$ in units of
standard candles (of the same $M$ and $R$). It measures the neutrino
cooling efficiency, the very important property of superdense matter
to be extracted from observations of cooling neutron stars. The
parameter $f_\ell$ should be taken at a temperature
$\widetilde{T}(t)$ of the cooling star. If $\ell(\widetilde{T})$ and
$\ell_\mathrm{SC}(\widetilde{T})$ have the same temperature
dependence, then $f_\ell$ is just a number, independent of
$\widetilde{T}$; otherwise it depends on $\widetilde{T}$, as discussed next.

The enhancement of $\ell(\widetilde{T})$ over $\ell_{\rm
SC}(\widetilde{T})$ can occur if more efficient neutrino emission
mechanisms are allowed in the inner cores of massive neutron stars.
These fast mechanisms lead to neutrino luminosity
$L_\nu^\infty(\widetilde{T}) \propto \widetilde{T}^6$. The strongest
enhancement over $L_{\nu \rm MU}^\infty$ (with $f_\ell \sim
10^6-10^7$ at $\widetilde{T} \sim 10^9$~K) can be provided by the
direct Urca processes in nucleon or nucleon/hyperon matter
(\citealt{lattimeretal91,prakashetal92}). They open at densities
higher than some threshold density $\rho_\mathrm{DU}$. Thus the
neutrino luminosity in non-superfluid low-mass neutron stars (where
the direct Urca process is forbidden) is our neutrino standard candle,
while at larger $M$ it is
much higher, leading to fast neutrino cooling. The density threshold
$\rho_{\rm DU}$ and the associated neutron-star mass threshold
$M_{\rm DU}$ for fast neutrino cooling are very model dependent
(see Sec.~\ref{sec:obs} and Fig.~\ref{fig:mr}).

Even if direct Urca processes are forbidden in the inner
neutron star core, the neutrino luminosity can still be enhanced
over $L_{\nu \rm MU}^\infty$ and can scale as
$L_\nu^\infty(\widetilde{T}) \propto \widetilde{T}^6$. This
happens provided the inner core has an exotic composition (contains
pion condensates, kaon condensates or free quarks) owing to
specific direct Urca-like processes in exotic matter. The neutrino
luminosity $L_\nu^\infty(\widetilde{T})$ depends on this
composition (e.g.,
\citealt{pethick92,yakovlevetal01,pageetal06,pageetal09}) but is
smaller than $L_{\nu \rm DU}^\infty(\widetilde{T})$, leading to a
smaller enhancement factor $f_\ell$. In pion condensed matter one
gets $f_\ell \sim 10^3-10^6$, while in kaon-condensed or quark
matter $f_\ell \sim 10^2-10^4$. Calculations of neutrino
emissivities and $\ell(\widetilde{T})$ for these cases are very
model dependent.

The situation is even more complicated in the presence of
superfluidity of baryons (Cooper pairing due to attractive
component of the baryon-baryon interaction) in neutron star cores.
In the simplest case of nucleon matter, one deals with
superfluidity of neutrons and protons. Hyperons and/or quarks (if
available) can also be in a superfluid state. Pion or kaon
condensates affect
superfluidity of baryons. Each
superfluidity is characterized by its own critical temperature
$T_c$ that depends on $\rho$. Calculations of $T_c(\rho)$ are
model dependent (e.g., \citealt{ls01}); various models give a
large scatter of $T_c(\rho)$ values ($\sim 10^8-10^{10}$~K or
higher). In central regions of massive stars, superfluidity
disappears [$T_c(\rho) \to 0$] because short-range nucleon-nucleon
repulsion destroys Cooper pairing.

Baryon superfluidity can greatly modify neutrino emission, and hence
affect $\ell(\widetilde{T})$ and $f_\ell$ (see, e.g.,
\citealt{yakovlevetal01,pageetal06,pageetal09}). First, strong
superfluidity ($T_c \gg T$) exponentially suppresses the traditional
neutrino reactions, which involve superfluid particles, and reduces
$\ell(\widetilde{T})$ because the gap in the energy spectra of
superfluid particles blocks the reactions. Second, superfluidity
initiates a different neutrino emission process due to Cooper pairing of
baryons, which increases $\ell(\widetilde{T})$.

Let us illustrate these statements by taking neutron stars with nucleon
cores as an example. Consider a low-mass star, which would cool via the
modified Urca process in a non-superfluid case. Strong (for
instance, proton) superfluidity in the core suppresses the modified
Urca process and reduces $\ell(\widetilde{T})$. If neutrons were
non-superfluid, the main neutrino emission would be provided by
neutrino pair production in neutron-neutron collisions
(neutron-neutron bremsstrahlung); $L_\nu^\infty(\widetilde{T})$
would be proportional to $\widetilde{T}^8$ (as for the modified
Urca process) but would be weaker than  $L_{\nu \rm
MU}^\infty(\widetilde{T})$; the star would cool slower than its
non-superfluid counterpart, leading to $f_\ell \sim 0.01-0.1$.
Now assume a mild triplet-state neutron superfluid in the core,
with maximum $T_c(\rho)$ of a few times of $10^8$~K. When the
temperature in the core drops below this maximum, Cooper pairing of
neutrons can initiate a sufficiently strong neutrino emission, with
$L_\nu^\infty(\widetilde{T}) \propto \widetilde{T}^8$. Such a star
would be a faster neutrino cooler than the standard candle (with
maximum $f_\ell \sim 10-100$). On the other hand, this extra
neutrino emission is negligible if the direct Urca process is open
(even slightly) in the core.

The effect of superfluidity in a massive star can be dramatic. The
direct Urca process can be formally allowed in its core but suppressed by
superfluidity. In this case, superfluidity transforms the fast
neutrino cooling into a slow one and effectively increases
$M_\mathrm{DU}$. It can greatly smooth out the transition from slow
to fast cooling with increasing $M$. However, very massive stars
would cool rapidly (if fast cooling is allowed by the EOS) because
superfluidity disappears
in their central regions (see above).

Recall that $\ell(\widetilde{T})$ depends also on the heat capacity
$C(\widetilde{T})$. In a non-superfluid star with a nucleon core,
$C(\widetilde{T})$ is mainly determined by neutrons (see above). The
proton contribution is $\sim$1/4 of the neutron one, and the
contribution of electrons and muons is $\sim$1/20 \citep{page93}.
Strong neutron superfluidity suppresses the neutron contribution,
and strong proton superfluidity suppresses the proton contribution,
but the electron and muon heat capacities always survive.

All in all, current theories predict different cooling rates
$\ell(\widetilde{T})$ which depend on many uncertain properties of
superdense matter. The temperature dependence $\ell(\widetilde{T})$
is mainly expected to be a power-law,
\begin{equation}
   \ell(\widetilde{T})= q \widetilde{T}^{n-1},
\label{eq:power-law}
\end{equation}
with $n=6$ for fast neutrino cooling and $n=8$ for slow cooling; $q$
determines the cooling efficiency. From (\ref{eq:power-law}) and
(\ref{eq:cool2}) at the neutrino cooling stage (after the initial
conditions are forgotten), one obtains the cooling relation
\begin{equation}
    \widetilde{T}(t)= [ (n-2)q t ]^{-1/(n-2)},
\label{eq:cool-pow}
\end{equation}
with
\begin{equation}
     \ell(\widetilde{T})= \widetilde{T}/[(n-2)t ].
\label{eq:ell}
\end{equation}
Such solutions have been presented in previous works many times
(see, e.g., \citealt{pethick92}). If one knows $t$, $n$ and
$\widetilde{T}$, then (\ref{eq:ell}) immediately gives the cooling
rate $\ell(\widetilde{T})$. For a fixed $t$ the rate is directly
proportional to $\widetilde{T}$. Eq.~(\ref{eq:cool-pow}) greatly
simplifies the solution of the cooling problem for a thermally
relaxed star
at the neutrino cooling stage. Note the weak dependence of
$\widetilde{T}$ on $t$ and on the cooling efficiency $q$, which
stems from the strong temperature dependence of the cooling rate.

It is instructive to introduce the parameter
\begin{equation}
   f_T={\widetilde{T}(t) \over \widetilde{T}_\mathrm{SC}(t)}=
   {n-2 \over n_\mathrm{SC}-2}\,
   {\ell(\widetilde{T})\over
   \ell_\mathrm{SC}(\widetilde{T}_\mathrm{SC})},
\label{eq:fc}
\end{equation}
where $n$ and $\widetilde{T}(t)$ refer to a given star, while
$n_\mathrm{SC}=8$ and $\widetilde{T}_\mathrm{SC}(t)$ refer to the
standard candle of the
same age, $M$ and $R$. $f_T$ measures the ratio of internal
temperatures of the star and the candle, as well as the ratio of
their cooling rates taken at their own internal temperatures.
Because of the strong
$\widetilde{T}$-dependence of the cooling rate $\ell$ [see
eq.~(\ref{eq:power-law})], the parameter $f_T$ [$\propto
1/\ell_\mathrm{SC}(\widetilde{T}_\mathrm{SC})$] is drastically
different from the ratio $f_\ell$ [$\propto
1/\ell_\mathrm{SC}(\widetilde{T})$; see eq.\ (\ref{eq:f-ell})].
According to eqs.~(\ref{eq:power-law}) and (\ref{eq:cool-pow}),
the cooling efficiency $f_\ell$ [calculated at a given stellar
temperature $\widetilde{T}(t)$] is related to $f_T$ by
\begin{equation}
    f_\ell={ q \over q_\mathrm{SC}}\, \widetilde{T}^{n-n_\mathrm{SC}}=
            {n_\mathrm{SC}-2 \over n-2} {1 \over
            f_T^{n_\mathrm{SC}-2}}.
\label{eq:f-ell1}
\end{equation}
For instance, if the star has $n=8$ and is twice as hot as the
corresponding standard candle should be ($f_T=2$), its cooling
efficiency is much lower than that of the candle
($f_\ell=1/2^{n_{\mathrm{SC}}-2}=1/64$). We will show
in Section~\ref{sec:method}
that $f_T$ can be easily constrained
from observations. However, it is $f_\ell$ that determines the
efficiency of neutrino emission and is the quantity of primary
importance.

We will also show that observations of the \casa\ are consistent
with $f_\ell \sim 0.01-100$. In this case it is reasonable to choose
$n=8$ (as we will do throughout the paper); $f_\ell$ and $f_T$ are
then independent of $t$ and $\widetilde{T}$. If in actuality
$n=6$,
while we have chosen $n=8$, then eq.~(\ref{eq:f-ell1})
underestimates $f_\ell$ by a factor of 1.5.

\subsection{Standard candles}
\label{sec:candle}
%

\begin{figure*}
 \includegraphics[width=\textwidth]{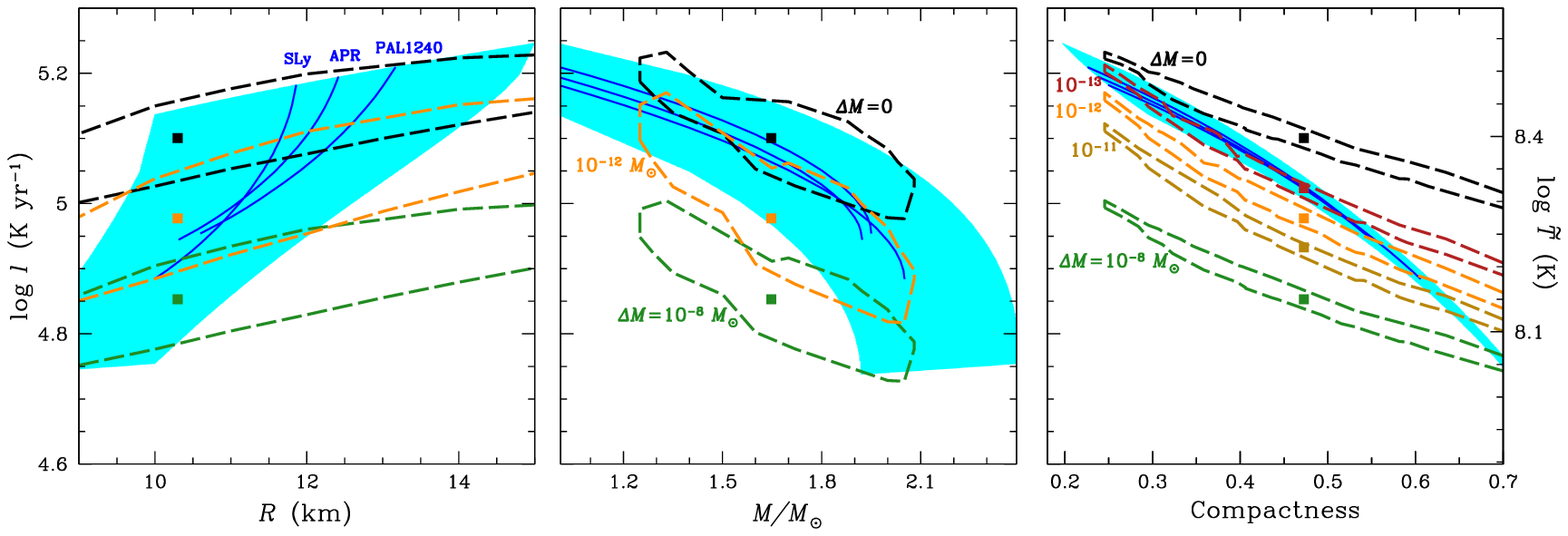}
\caption{(Color on line) Cooling rates $\ell$
[defined by eq.~(\ref{eq:cool2})]
inferred from the observations of \casa\ at 99\% confidence level
(dashed lines) as a function of radius (left), mass (middle) and
compactness (right). The rates are inferred assuming either an
iron heat-blanketing envelope ($\Delta M=0$) or
envelopes with outer carbon shells
 of mass $\Delta M = 10^{-12}M_\odot$ and
$10^{-8}M_\odot$. On the right panel we add the dashed lines for $\Delta M
= 10^{-13}$ and $10^{-11}\,M_\odot$. The shaded regions are
theoretical $\ell_\mathrm{SC}$ values for standard candles with $M$
and $R$ from the shaded region in Fig.~\ref{fig:mr}.  Solid lines
refer to standard candles using the SLy, APR and PAL1240
EOSs. The right vertical scale on the right panel shows the
internal temperature of the \casa.
\label{fig:coolfun} }
\end{figure*}

We simulate cooling of many standard candles, taking 17 nucleonic
EOSs that describe the neutron star core. The EOSs include all 9
original versions of the phenomenological PAL EOS
\citep{prakashetal88}, as well as 3 other versions of this EOS
with the symmetry energy of nuclear matter proposed by
\citet{pageapplegate92}; the APR EOS \citep{akmaletal98}; the SLy
EOS \citep{douchinhaensel01}; and three new modifications (i--iii)
of the PAL EOS. These modifications correspond to (i)
$K_0=400$~MeV and model 2 for the symmetry energy from
\citet{prakashetal88}; (ii) $K_0=300$~MeV and model 3; and (iii)
$K_0=400$~MeV and model 3; the parameter $\sigma$ in
\citet{prakashetal88} was taken to be zero in all three cases.
These three EOSs are definitely superstiff and unrealistic. In
particular, they predict the existence of very massive stars (with
maximum masses up to $\sim 2.8\, M_\odot$). We include them to
check that the scaling relations presented below can be
extrapolated to supermassive neutron stars. For any chosen EOS, we
consider neutron star models of different $M$, from $1\,M_\odot$
to the maximum mass, with a step size of 0.1\,$M_\odot$ or
smaller. For those models where the direct Urca process is
allowed, we switch the process off to ensure that we consider only
standard candles. In this way, we take about 170 neutron star
models which cover a large area of the mass-radius diagram, with
$R$ from about 9 to 15 km. We simulate cooling of all our models
and determine $\widetilde{T}(t)$ and $T_\mathrm{s}(t)$. In the
simulations, we assume non-magnetic iron heat blanketing
envelopes. However, the values of $\widetilde{T}(t)$ are almost
insensitive to the composition of the envelopes (which determine
$T_\mathrm{s}$) at the neutrino cooling stage (see
Section~\ref{sec:relaxed}).

We find that the internal temperature
$\widetilde{T}(t)=\widetilde{T}_{\rm SC}(t)$ of all thermally
relaxed neutron stars (standard candles) at the neutrino cooling
stage can be approximated by
\begin{equation}
    \widetilde{T}_{\rm SC}(t)=3.45\times 10^8\mbox{ K }(1-x)\, \left[1+
    0.12\, \left(R \over 10~{\rm km}\right)^{2} \right]
    \, \left(t_C \over t \right)^{1/6},
\label{eq:scaling}
\end{equation}
where the compactness $x$ is given by (\ref{eq:x})
and $t_C$ is some fiducial (normalization) timescale. The latter
is set equal to the age of the Cas~A  supernova remnant ($t_C=330$
yr) that is convenient for our purpose. The time dependence of
$\widetilde{T}_{\rm SC}$ in (\ref{eq:scaling}) is taken in
accordance with eq.\ (\ref{eq:cool-pow}). Calculations show that
this time dependence is, indeed, accurate at the neutrino cooling
stage (see Section \ref{sec:casa}). It first establishes in the
core which thermally equilibrates faster than the inner crust.
After the global internal thermal relaxation is over, it becomes
accurate in the entire star (except for the heat blanketing
envelope). The numerical factors 3.45 and 0.12 in (\ref{eq:scaling})
are obtained by fitting the calculated values
$\widetilde{T}(t_C)$. The accuracy of the fit is sufficiently
good. The root mean square relative fit error is $\approx 3\%$
(which translates into the 1.5\% error of $T_\mathrm{s}$ because
$T_\mathrm{s} \propto T_\mathrm{b}^{1/2}$); the maximum error over
all models is $\approx 10\%$ for an unrealistic (supermassive)
model with $M=2.85\,M_\odot$; the maximum error over $\sim$160
models with $M<2.4\,M_\odot$ is $\approx 7\%$ (for the
1.9$M_\odot$ model with the SLy EOS).

Note, that our cooling curves contain systematic uncertainties
which come from uncertainties in the calculations of the neutrino
luminosity $L_{\nu \rm MU}^\infty$ and heat capacity $C$ of the
core. The main uncertainties are those in the effective masses of
neutrons and protons, $m_\mathrm{n}^*$ and $m_\mathrm{p}^*$, and
in the squared matrix element $|{\cal M}|^2$ of the modified Urca
process in the core. These quantities should be calculated
self-consistently for a specific EOS (using specific many-body
theory) as a function of density $\rho$, but such calculations
(which would be most desired!) are currently unavailable. Instead,
we set somewhat arbitrarily
$m_\mathrm{n}^*=m_\mathrm{p}^*=0.7 m_\mathrm{n}$ ($m_\mathrm{n}$
being the free neutron mass), and we use the standard
approximate expression for $|{\cal M}|^2$ from
\citet{yakovlevetal01}. To estimate the above uncertainties, let
us introduce a typical nucleon effective mass $m_\mathrm{N}^*$ in
dense matter and note that $L_{\nu \rm MU}^\infty \propto
(m_\mathrm{N}^*)^4|{\cal M}|^2$ and $C(T) \propto m_\mathrm{N}^*$.
Then from eq. (\ref{eq:cool-pow}) we conclude that $\widetilde{T}
\propto 1/q^{1/6} \propto (m_\mathrm{N}^*)^{-1/2} \,|{\cal
M}|^{-1/3}$. For instance, allowing a 20\%  uncertainty in
$m_\mathrm{N}^*$ and 30\% (uncorrelated) uncertainty in $|{\cal
M}|^2$, we obtain $\sim 15$\% uncertainty in $\widetilde{T}$ and
7\% in $T_\mathrm{s}$. We will not take these uncertainties into
account in our further analysis (which is a common practice in the
neutron star cooling theory) but warn about them. Taking the same
nucleon effective mass $m^*_\mathrm{N}$ for neutrons and protons
would be inaccurate in advanced simulations but is sufficient for
estimating the uncertainties of $\widetilde{T}$.

Eq.\ (\ref{eq:scaling}) gives a very simple method to calculate the
standard slow cooling of thermally relaxed neutron star of any
reasonable mass and radius during the neutrino stage. It does not
require detailed knowledge of the internal stellar structure and
microphysics of superdense matter. Because $\widetilde{T}$ is
independent of the properties of the heat blanketing envelope at the
neutrino cooling stage, eq.\ (\ref{eq:scaling}) is not biased by
these properties but allows one to take them into account while
calculating $T_\mathrm{s}$. One can easily determine $\widetilde{T}$
for given values of $t$, $M$ and $R$, and find
$T_\mathrm{b}=\widetilde{T}/\sqrt{1-x}$ at the bottom of the heat
blanketing envelope. Then it is a simple task to calculate
$T_\mathrm{s}$ using the $T_\mathrm{s}(T_\mathrm{b})$ relation.

For example, the shaded regions in Fig.~\ref{fig:ts} represent the
effective surface temperatures ($T_\mathrm{s}$ and
$T_\mathrm{s}^\infty$) of standard candles of age $t=330$ yr. Their
masses and radii are taken from the (realistic EOS) shaded region in
Fig.~\ref{fig:mr}. All of them assume an iron heat
blanketing envelope; the exception is for the two lightly-shaded regions
in the right panel -- these are for heat blanketing envelopes
containing $\Delta M = 10^{-8}M_\odot$ of carbon. Note a striking
feature of theoretical standard candle curves -- they are described
by a well-defined (nearly universal) function of compactness $x$,
but not of $M$ and $R$. This is a remarkable manifestation of
General Relativity. The presence of the maximum amount of carbon
$\Delta M \sim 10^{-8}M_\odot$ in the heat blanketing envelope does
not violate the existence of a well-defined dependence of
$T_\mathrm{s}$ and $T_\mathrm{s}^\infty$ on $x$, but raises the
effective temperatures by $\approx 25\%$ for standard candles
of the \casa\ age.

At $x \lesssim 0.5$ the redshifted surface temperature of all
standard candles with iron blanketing envelopes is nearly the same,
$T_\mathrm{s}^\infty \approx 1.4$ MK. With increasing $x$, the
redshifted temperature $T_\mathrm{s}^\infty$ decreases, down to
$\approx 1.1$ MK for standard candles with the maximum compactness
$x \approx 0.7$; in contrast, the non-redshifted temperature
$T_\mathrm{s}$ increases. Thus compact standard candles would appear
colder for a distant observer but would be hotter for a local
observer. The differing behavior of $T_\mathrm{s}^\infty$ and
$T_\mathrm{s}$ is another manifestation of General Relativity.
The weak sensitivity of $T_\mathrm{s}^\infty(t)$ to variations of $M$ and $R$
for standard candles was first noted by
\citet{pageapplegate92} and later discussed in the literature (e.g.,
\citealt{yakovlevpethick04}, and references therein); here we
quantify this effect.

The three solid lines in Fig.\ \ref{fig:ts} show
$T_\mathrm{s}$ and $T_\mathrm{s}^\infty$ for standard candles using
the SLy, APR and PAL1240 EOSs (their $M-R$ relations are displayed
in Fig.~\ref{fig:mr}). We plot these curves using eq.\
(\ref{eq:scaling}) and assuming iron heat blanketing envelopes.
These curves are similar to each other, not only as a function of $x$,
but also as a function of $M$ (because the three EOSs are alike).

Theoretical cooling rates [eq.~(\ref{eq:power-law}), along with
eq.~(\ref{eq:scaling})] for standard candles of the \casa\ age are
shown by the shaded regions in Fig.~\ref{fig:coolfun}. The three panels
give these rates versus $R$, $M$ and $x$. The masses and radii of
neutron stars are again taken from the shaded region in Fig.\
\ref{fig:mr} and the solid lines are again the theoretical rates for
standard candles with the SLy, APR and PAL1240 EOSs. All these rates
are independent of the properties of the blanketing envelopes.
We see again that there is a well-defined relation between the
cooling rate $\ell$ and compactness $x$.

Since $\ell(\widetilde{T})$ unambiguously determines $\widetilde{T}$
(for $n=8$), the right vertical scale in Fig.~\ref{fig:coolfun}
can be used to calibrate the internal neutron star temperature
$\widetilde{T}$. The expected values of $\widetilde{T}$ are a few
$\times 10^8$~K. The non-redshifted local internal temperature $T$
is somewhat higher. The relation between $\widetilde{T}$ and $T$ is
similar to that between $T_\mathrm{s}^\infty$ and $T_\mathrm{s}$ in
Fig.\ \ref{fig:ts}. The difference between $T$ and $\widetilde{T}$
in the neutron star core is larger than in the crust because of
stronger gravitational redshifts in the core. The core is hotter
($T$ is higher) than the crust for a given isothermal interior
(given $\widetilde{T}$).

\subsection{Determining the cooling rate}
\label{sec:method}
%

We now propose a simple and robust method to determine the cooling
rate $\ell(\widetilde{T})$ in a thermally relaxed star at the
neutrino cooling stage. Assuming some values of $M$ and $R$, a model
for the heat blanketing envelope, and taking a measured
(constrained) value $T_\mathrm{s}$ (or $T_\mathrm{s}^\infty$), we
can find (constrain) $\widetilde{T}$. With a given (or assumed) age
$t$, the cooling rate $\ell$ can then be determined (estimated) from
(\ref{eq:ell}) (with $n=8$ as discussed above).

The application of this method to the \casa\ is presented in
Fig.~\ref{fig:coolfun}, which shows values of $\ell$ (left vertical
scale) and $\widetilde{T}$ (right vertical scale) of the \casa\
versus $R$, $M$ and $x$. These values are inferred from spectral
fits of the \casa\ at the 99\% confidence level (Fig.~\ref{fig:ts}).
The accuracy of inferring $\widetilde{T}$ is limited by
the accuracy of our models for heat blanketing envelopes; the
accuracy of inferring $\ell$ is also affected by the assumption that
$n=8$. The upper (dashed) lines of $\ell$ and $\widetilde{T}$ are obtained
using the neutron star models with iron heat blankets. Lower
lines are for models with carbon shells of different mass $\Delta
M$. We again see that the values of $\ell$ and $\widetilde{T}$ are
well-defined functions of $x$ (for a given $\Delta M$). The inferred
values of $\ell$ and $\widetilde{T}$ of the \casa\ for realistic
values of $M$, $R$, $x$ and $\Delta M$ vary within the factor $f_T
\sim 3$. The advantage of Fig.~\ref{fig:coolfun} is that it gives
absolute values of the cooling rate $\ell$, an important parameter
of the \casa.

\begin{figure*}
 \includegraphics[width=\textwidth]{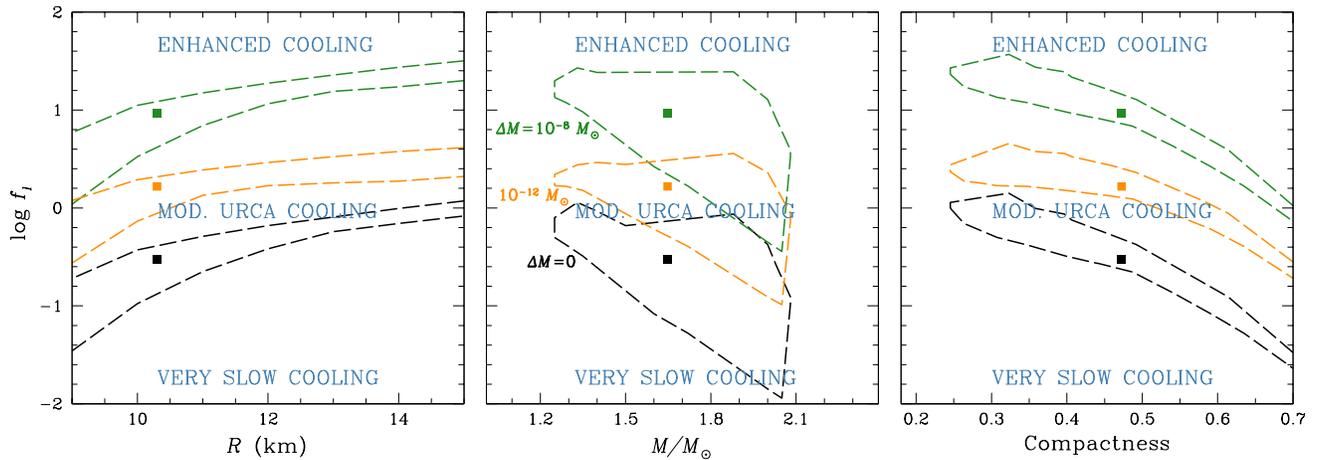}
\caption{(Color on line) Contours of \casa\ cooling efficiency
$f_\ell$
[defined by eq.~(\ref{eq:f-ell})] as a function of radius (left),
mass (middle) and compactness (right) at the 99\% confidence
level. The heat-blanketing envelopes are composed of either iron
($\Delta M=0$) or carbon shells of mass $\Delta M =
10^{-12}M_\odot$ or $10^{-8}M_\odot$.
\label{fig:fl} }
\end{figure*}

We now take the inferred values of
$\widetilde{T}$, use eq.~(\ref{eq:f-ell}) and calculate the
cooling efficiency $f_\ell$, which is another important parameter
of the \casa\ cooling. As discussed in
Section~\ref{sec:neutrinostage}, relatively small variations of
$\widetilde{T}$ yield
much larger variations of  $f_\ell$. This is demonstrated in Fig.~\ref{fig:fl},
which plots $f_\ell$ as a function of $R$, $M$ and
$x$. Lower, middle and upper lines refer to the blanketing
envelopes  with $\Delta M/M_\odot=0,~10^{-12}$ and $10^{-8}$,
respectively. The figure directly compares the \casa\ neutrino
cooling rate to that of the standard candle.

Figures~\ref{fig:ts}--\ref{fig:fl} impose constraints on the \casa\
cooling rate. Recall that we can specify the rate by the single
parameter $f_\ell$ (provided we fix $n=8$). Our results suggest that
$f_\ell$ (as well as $\ell$, $T_\mathrm{s}$, $T_\mathrm{s}^\infty$)
is a nearly universal function of $x$ and $\Delta M$.
If the \casa\ has low compactness $x \lesssim 0.4-0.5$ and possesses
either an iron blanketing envelope ($\Delta M=0$) or an envelope
with a very low mass carbon shell ($\Delta M \lesssim
10^{-13}\,M_\odot$), we find $f_\ell \sim 1$ (i.e., standard candle
cooling).  For larger $x$ but the same heat blanket, we obtain lower
$f_\ell$ (very slow cooling). For the maximum compactness $x \approx
0.7$, we find $f_\ell \sim 0.02-0.03$.
If the \casa\ has a massive carbon heat blanket, then a higher cooling
efficiency is required. With the most massive carbon shell $\Delta M
\sim 10^{-8} M_\odot$ at $x \lesssim 0.4-0.5$, the \casa\ cools
faster than the standard candle, with $f_\ell \sim 30-50$. However,
$f_\ell$ decreases with increasing $x$ and reaches the standard
level at $x \approx 0.7$.
Note the opposite $\Delta M$-dependence of the cooling rate $\ell$
(Fig.~\ref{fig:coolfun}) and cooling efficiency $f_\ell$
(Fig.~\ref{fig:fl}). The star with smaller $f_\ell$ cools slower and has
higher temperature (and hence higher $\ell$) than the standard
candle of the same age [see eq.~(\ref{eq:f-ell1})].

We thus set robust restrictions on the cooling efficiency
$f_\ell$ of the \casa\ as a thermally relaxed cooling neutron star.
Unfortunately, the values of $f_\ell$ do not specify unambiguously
the neutrino emission mechanisms within the star (Section
\ref{sec:neutrinostage}) and can be realized by different physical
models for the interior of \casa. For instance, $f_\ell \sim 1$ is
consistent with the modified Urca process in a non-superfluid
neutron star core. It is equally consistent with enhanced
neutrino emission that is partially suppressed by superfluidity.
Thus there is ambiguity in the theoretical interpretation.
We can explain the \casa\ data assuming different $x$ and
$\Delta M$, while the values of $f_\ell$, inferred at fixed $x$
and $\Delta M$, can be consistent with many models of \casa\
interiors.

\section{\casa\ among other cooling stars}
\label{sec:casa}
%

\begin{figure}
\includegraphics[width=\columnwidth,bb= 15 145 545 650]{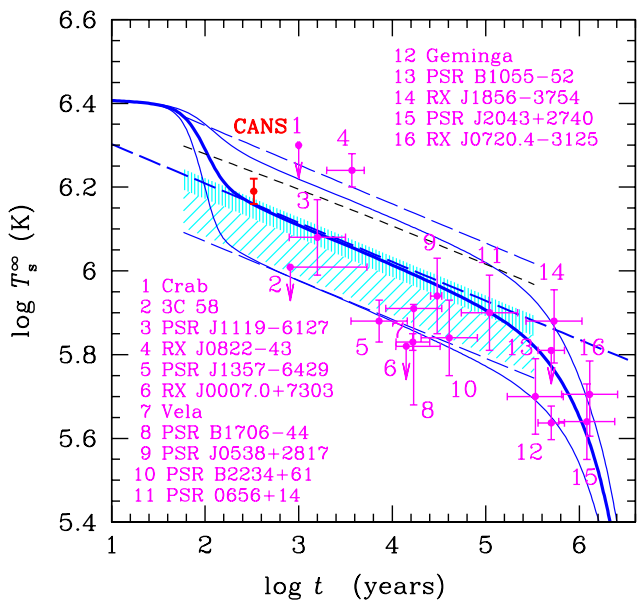}
\caption{ (Color on line) Theoretical cooling curves
$T_\mathrm{s}^\infty(t)$ compared with observations of cooling
isolated neutron stars. The thick solid line is the standard candle
(the $1.4 M_\odot$ APR star) with an iron heat blanketing envelope.
Upper and lower thin solid lines are for the same star but the
neutrino emission rate is artificially reduced and enhanced,
respectively, by a factor of 30. The thick long-dashed line is the
same as the thick solid line but calculated from the analytic
expression (\ref{eq:scaling}) that is valid for thermally relaxed
stars at the neutrino cooling stage. The densely shaded strip is
covered by numerous cooling curves of standard candles having iron
envelopes and $x<0.5$; lightly-shaded strip is the same but for
standard candles with $0.5 \leq x <0.7$. The upper and lower thin
long-dashed lines are analytic approximations (\ref{eq:scaling}) of the
corresponding thin solid curves. The thin short-dashed line is the
analytic approximation of the cooling curve for the $1.4M_\odot$ APR
star with a blanketing envelope containing $\Delta
M=10^{-8}M_\odot$ of carbon.
\casa\ is the data point at
$t \approx 330$~yr,
  while the values
for the other data points are taken from
references,
cited
in \citet{yakovlevetal08,kaminkeretal09}.
\label{fig:cool} }
\end{figure}

Figure~\ref{fig:cool} plots the \casa\ $T_\mathrm{s}^\infty$
limits on the $T_\mathrm{s}^\infty-t$ plane, as well as the data
on other cooling isolated neutron stars,
The \casa\ limits are those obtained from 99\% confidence contours
(Fig.\ \ref{fig:ts}).
 Observational data on other sources are
taken from references cited in
\citet{yakovlevetal08,kaminkeretal09}. The \casa\ is the youngest
in the family of cooling neutron stars whose surface temperatures
are measured (constrained) more or less reliably.

The thick solid line in Fig.~\ref{fig:cool} demonstrates the
cooling of a standard candle. For this case, we take a neutron
star with $M=1.4 M_\odot$,
$R=12.14$ km ($x=0.34$),
the APR EOS in the core and a heat-blanketing envelope composed of
iron. The initial part of the cooling curve is almost flat, which
indicates that the stellar interior is not thermally relaxed
\citep{lattimeretal94,gnedinetal01}. The rapid drop of the surface
temperature signals the end of thermal relaxation. By the current
age of the \casa\ ($\approx 330$~yr), the star has an isothermal
interior, and the results of Section \ref{sec:stage2} apply. The
curve is reasonably consistent with the \casa\ observations, in
agreement with the results of Section \ref{sec:method}.

The thick long-dashed cooling curve is calculated using the
approximate cooling relation given by eq.~(\ref{eq:scaling}). It
is valid during the neutrino cooling stage in a thermally relaxed
star. It accurately describes the cooling at $t \gtrsim 250$ yr
(when the thermal relaxation is over) until $t \lesssim 10^5$ yr
(while photon cooling is still unimportant). The densely-shaded
region around this cooling curve is also calculated from eq.\
(\ref{eq:scaling}). We only show it in the limited range of $t$
because eq.\ (\ref{eq:scaling}) is inapplicable for very young and
very old stars. This region is covered by cooling curves of
standard candles (with iron envelopes); masses and radii are taken
from that part of the central shaded region in Fig.~\ref{fig:mr}
where $x < 0.5$. These cooling curves nearly coincide and are
almost consistent with the \casa\ data (in agreement with the
results of Section \ref{sec:method}; see Figs.~\ref{fig:ts} and
\ref{fig:coolfun}). The lower lightly-shaded region in
Fig.~\ref{fig:cool} is covered by similar cooling curves of more
compact standard candles ($0.5 \leq x < 0.7$) from the shaded
region in Fig.~\ref{fig:mr}. With increasing compactness, standard
candles have lower $T_\mathrm{s}^\infty$ and cannot explain the
\casa\ data. Neutrino emission would need to be reduced to be
consistent with the observations. A reduction of the cooling rate
to $f_\ell \sim 0.02-0.03$ is required for the maximum compactness
$x\approx 0.7$.

The thin solid lines in Fig.~\ref{fig:cool} are again the cooling
curves of the 1.4$M_\odot$ APR star (with the iron heat
blanketing envelope). However, in this case the efficiency of
neutrino emission is artificially enhanced (the lower curve) or
suppressed (the upper curve) by a factor of 30 (i.e., $f_\ell =
30$ or 1/30, respectively). The thin dashed lines are analytic
approximations of these curves; they are calculated from
eq.~(\ref{eq:scaling}), in which we introduce the factor
$f_\ell^{-1/6}$ in accordance with eq.~(\ref{eq:cool-pow}) to
renormalize the efficiency of neutrino emission. There is a good
agreement between the lower thin solid and dashed curves (when the star
is thermally relaxed and cools via the neutrino emission). Also note that the
1.4$M_\odot$ APR star with $f_\ell=30$ cools almost as a very
compact ($x=0.7$) standard candle ($f_\ell=1$). On the other hand, we
obtain a slight disagreement between the upper solid and long-dashed
curves (for our basic APR star with strongly reduced neutrino
emission, $f_\ell=1/30$). This is because the
cooling of
neutron stars with very low neutrino emission from the core can be
affected by neutrino emission from the crust. When we artificially
reduce the emission from the crust, the upper thin solid curve
shifts to the upper thin long-dashed curve (the shifted curve is not
shown in Fig.~\ref{fig:cool}). We have checked that this effect
becomes noticeable only in neutron stars with very low neutrino
cooling rate ($f_l \lesssim 0.03$). Similar effects were discussed
by \citet{kaminkeretal01} who studied very slowly cooling neutron
stars with the neutrino emission in the core greatly suppressed by
strong nucleon superfluidity. Thus our analytic approximation
becomes inaccurate for $f_\ell \ll 0.03$. One should also be careful
in applying this approximation to the case of $f_\ell \gg 30$
because it is reasonable to expect that the cooling rate index $n$
can change from $n=n_\mathrm{SC}=8$ to $n=6$ (see
Section~\ref{sec:neutrinostage}).

Finally, the thin short-dashed line in Fig.~\ref{fig:cool} is the
analytic cooling curve of the APR $1.4 M_\odot$ standard candle with
a (maximal $\Delta M=10^{-8}M_\odot$) carbon heat blanketing envelope.
Recall that
during the neutrino cooling stage such a star has
the same internal temperature as a star with an iron envelope;
nevertheless, it has a hotter surface because the carbon envelope is
more heat transparent. The thin short-dashed curve is close to the
upper solid and long-dashed curves meaning that the surface
temperature of a standard candle with the carbon envelope is
nearly the same as that of a star with lower cooling rate ($f_\ell
\sim 0.03$) and iron envelope. If the \casa\ has a carbon envelope
and is not too compact ($x \lesssim 0.5$), then its neutrino cooling
rate should be over that of the standard candle by a factor of
30--50 (in agreement with the results of Section \ref{sec:method}).
As the compactness increases, the surface temperature of the standard
candle with a carbon envelope will tend to the \casa\ range;
for the highest compactness and a carbon envelope, \casa\ would
cool as a standard candle.

An analysis of current theories  of isolated neutron stars is
given, e.g., by
\citet{yakovlevpethick04,pageetal06,yakovlevetal08,pageetal09}. We
note that the
surface temperatures of
nearly all
stars
lie
between the upper and lower thin curves in Fig.\ \ref{fig:cool}
(the observed isolated stars do not require extremely slow or fast
neutrino cooling). Our results indicate that the coldest of them
(like the Vela pulsar), which have usually been treated as rapid
neutrino coolers, can in fact be very compact standard candles.
Note that the surface temperatures of many isolated neutron stars
(especially the colder ones) can also depend on strong magnetic
fields, which affects heat transport in the blanketing envelopes
\citep[see][]{potekhinetal03,potekhinetal07}.

\section{Is \casa\ relaxed?}
 \label{sec:coolnrex}

As shown in Section \ref{sec:stage2}, observations of the \casa\ are
consistent with the assumption that the star is thermally relaxed
and cools relatively slowly (with a cooling efficiency $f_\ell \sim
0.02-50$). In this section, we examine the possibility that the
\casa\ is cooling much faster but its surface is not too cold
because it is not yet thermally relaxed. We stress that cooling
theories with the standard microphysics
of neutron stars give relaxation
times shorter than 300 yr (e.g., \citealt{lattimeretal94,gnedinetal01}).
This means that
the possibility considered here requires non-standard cooling scenarios
which delay relaxation.

In a young neutron star, the crust is hotter than the core because
of lower neutrino emission in the crust. Relaxation implies thermal
equilibration of the crust with the core that is mainly regulated by
the physics of the crust
\citep{lattimeretal94,gnedinetal01,shterninyakovlev08}. It can be
delayed by lowering the thermal conductivity and neutrino luminosity
in the crust and by considering neutron star models with a thicker
crust (which increases the thermal diffusion timescale in the crust).

\begin{figure}
\begin{center}
 \includegraphics[width=0.85\columnwidth]{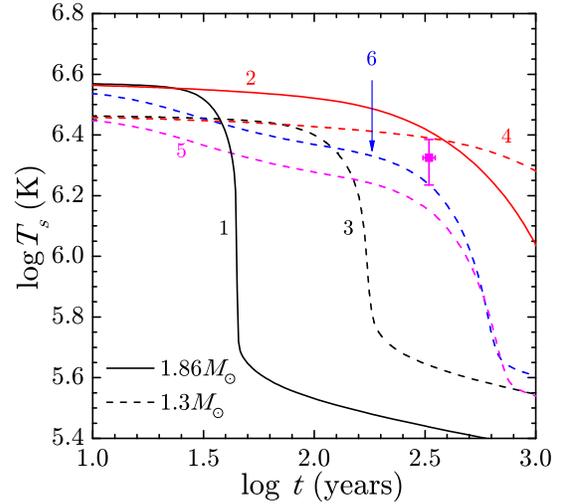}
\end{center}
\caption{(Color on line) Cooling curves calculated for $1.86
M_\odot$ neutron star (solid lines) and $1.3 M_\odot$ star (dashed
lines) with the PAL1240 EOS compared with observations of the \casa.
Curves~1 and 3 correspond to standard physics input. Curves~2 and 4
employ the lowest limit on the thermal conductivity in the crust.
Curve~5 is the same as 4 but adding moderate neutron superfluidity
in the crust. Curve~6 is the same as 5 but the star has most massive
carbon shell in the heat blanketing envelope. \label{fig:coolnrex} }
\end{figure}

Figure~\ref{fig:coolnrex} gives a number of cooling curves
$T_\mathrm{s}(t)$ which are calculated for two neutron star models
based on the PAL1240 EOS. We assume that the neutron stars have
non-superfluid cores (since we focus on delaying relaxation by
tuning the physics of the crust). The first model  has
$M=1.86M_\odot$ and $R=12.9$~km. Recall that the direct Urca process
is allowed for this EOS at $M_\mathrm{DU}=1.25 M_\odot$. Here we do
not switch this process off (as in the calculations presented in
Section~\ref{sec:stage2}). As a result, the neutrino cooling rate is
much higher [with $f_\ell \sim 10^6$; see eq.~(\ref{eq:f-ell})] than
for standard candles. The second neutron star model has
$M=1.3M_\odot$ and $R=12.9$~km. Its mass again exceeds
$M_\mathrm{DU}$, and the neutrino cooling rate is huge. This star
has a thicker crust than the $1.86M_\odot$ star,
and its relaxation time can be longer.  The displayed \casa\
errorbars are estimates from Fig.~\ref{fig:ts}.

Curve~1 in Fig.~\ref{fig:coolnrex} shows the cooling of the $1.86
M_\odot$ star with the standard physics of the core and crust. The
end of the thermal relaxation manifests itself in the drop of
$T_\mathrm{s}(t)$ at $t \sim 40$ yr. By the current age of \casa,
the star is thermally relaxed; it is much colder than the \casa\ (in
agreement with the results of Section \ref{sec:stage2}). The largest
delay of relaxation can be achieved by lowering the thermal
conductivity in the crust. We demonstrate this by using a model
thermal conductivity proposed by \citet{brown2000} (curve~2 in
Fig.~\ref{fig:coolnrex}). This model corresponds to an amorphous
crust and can be regarded as the lowest limit on the crustal
conductivity. The relaxation time is now longer than the \casa\ age.
By taking the thermal conductivity to be lower than the normal one
but above the lower limit, we can match any point in the region
between curves~1 and 2 and thus explain the \casa\ data. However,
the hypothesis of low conductivity contradicts observations of
thermal relaxation in quasi-persistent X-ray transients
\citep[see][and references
therein]{shterninetal2007,browncumming2009}.

Curve~3 shows the cooling of the $1.3 M_\odot$ neutron star (with a
thicker crust) assuming standard physics in the crust. The
relaxation time is longer than for the $1.86 M_\odot$ star (curve~1),
but the delay is insufficient to explain the \casa\ observations.
Curve~4 demonstrates the cooling of the same $1.3 M_\odot$ star but
with the lowest crustal thermal conductivity \citep{brown2000},
corresponding to curve~2. In this case, relaxation is rather slow.
Again by taking intermediate values of thermal conductivity, we can
fit any point between curves~3 and~4.

Relaxation can be accelerated if the neutron heat capacity in
the inner crust is suppressed by neutron superfluidity. This
effect is illustrated by curve~5, which is calculated using the
lowest thermal conductivity but assuming moderate neutron
superfluidity in the crust (c.f., curve~4). The relaxation time
drops to 700~yr, about twice the age of \casa.
Finally, curve~6 is the same as 5 but for the star with the carbon
heat blanketing envelope ($\Delta M= 2 \times 10^{-8} M_\odot$).
This star is the same inside but looks hotter
from the outside and has the same relaxation time.

Thus the observations of the \casa\ can be compatible with a
neutron star that undergoes fast cooling (via the direct Urca
process) but is still not thermally relaxed. However, this can
only be done by employing a very low thermal conductivity
throughout the neutron star crust,
which seems unrealistic.

\section{Conclusions} \label{sec:concl}

We have used the cooling theory of isolated neutron stars and
suggested a robust method (Section \ref{sec:method}) to infer the
neutrino cooling rate $\ell$ [defined by eq.\ (\ref{eq:cool2})] of
stars from observations of their thermal radiation at the neutrino
cooling stage (ages $t \lesssim 10^5-10^6$ yr) after the end of
their internal thermal relaxation. We simulated cooling of many
models (wide ranges of EOSs, masses and radii) of standard candles
(neutron stars with non-superfluid nucleon cores which cool via
the modified Urca process) and formulated simple relations
(Section \ref{sec:candle}) which determine the cooling rates and
cooling dynamics during the neutrino cooling stage. We have shown
that ultra-compact standard candles with compactness $x \sim 0.7$
have noticeably lower redshifted surface temperatures
$T_\mathrm{s}^\infty$ than their less compact counterparts ($x
\lesssim 0.5$). Based on these results, we developed a method to
compare the inferred cooling rates to the rates of standard
candles and to determine the neutrino cooling efficiency $f_\ell$
[given by
eq.~(\ref{eq:f-ell})] of observed neutron stars. We have
shown (Section \ref{sec:stage2}) that many physical properties,
which characterize the thermal state of a cooling neutron star
(such as the surface and internal temperatures, cooling rate), are
well-defined functions of the stellar compactness $x$, rather than
the stellar mass $M$ or radius $R$. Self-similarity properties in
the cooling of standard candles [eq.~(\ref{eq:scaling})], their
universal dependence on the compactness $x$, and the relatively
low $T_\mathrm{s}^\infty$ of ultra-compact standard candles have
been previously overlooked in the cooling theory. The suggested
method is practical. Fundamental uncertainties in the physics of
superdense matter in neutron stars are incorporated into the
parameter $f_\ell$ (or the cooling rate $\ell$) which can be
determined (constrained) from observations.

We applied our theoretical formalism to study the thermal state of
the neutron star in the Cas~A supernova remnant. We base our study
on the recent results of \citet{hoheinke09}, which showed that the
observed X-ray radiation from this neutron star can be interpreted
as thermal radiation from a carbon atmosphere (and the radiation is
emitted from the entire stellar surface). We discussed the data
(Section \ref{sec:obs}) and provided a theoretical interpretation
(Section \ref{sec:stage2}), assuming that the Cas~A neutron star (\casa) is
thermally relaxed. Our analysis shows that the
\casa\ cooling rate
 is $\ell \sim 10^5$ K~yr$^{-1}$ and its internal
temperature $\widetilde{T} \sim (2-3) \times 10^8$~K
(Fig.~\ref{fig:coolfun}). Its thermal state is mainly determined by
the cooling efficiency $f_\ell$, compactness $x$ and mass $\Delta M$
of light elements in the heat-blanketing envelope. The magnetic
field $B \lesssim 10^{11}$~G expected in the surface layers of the
\casa\
\citep{gotthelfhalpern07,gotthelfhalpern09,hoheinke09,halperngotthelf10}
does not affect its cooling.

If the \casa\ has a moderate compactness $x \lesssim 0.5$ and
possesses an iron heat blanketing envelope or an envelope with a
low mass shell of carbon ($\Delta M \lesssim 10^{-13}\,M_\odot$),
we find $f_\ell \sim 1$ (implying standard candle cooling). If it
has the same heat blanket but larger $x$ we obtain a lower $f_\ell$
(slower neutrino cooling), with a minimum $f_\ell \sim 0.02-0.03$
for $x_\mathrm{max} \approx 0.7$. For the most massive carbon shell
($\Delta M \sim 10^{-8} M_\odot$) and moderate $x$, the \casa\ would
have $f_\ell \sim 30-50$ (implying a cooling rate
enhanced with respect to the standard candle), but for the same
shell and maximum $x$ it would have the standard-candle cooling
efficiency $f_\ell \sim 1$. In summary, we find $0.02 \lesssim
f_\ell\lesssim 50$ for the \casa. Though these results do not give
an unambiguous physical picture of the thermal structure of the
\casa, they do impose firm and robust constraints.

We also examined the possibility that the \casa\ has a high neutrino
cooling rate but is still not thermally relaxed and, therefore, is
not very cold (Section \ref{sec:coolnrex}). We found that this is
possible (though not likely)
provided the neutron star has a very low thermal conductivity in
the crust.

Note that we considered heat blanketing envelopes made of iron, as well as
envelopes composed of an outer carbon shell. We could have
complicated the model, e.g., introducing an oxygen shell beneath the
carbon shell, but we do not expect qualitatively different results.

It would be instructive to apply the above analysis to other
cooling neutron stars, in an attempt to reconstruct the cooling
rate function $\ell$. It is expected to be one and the same
function for all neutron stars, and it contains
an important information on neutron star structure.

\section*{acknowledgements}

WCGH appreciates the use of the computer facilities at the Kavli
Institute for Particle Astrophysics and Cosmology. DGY, PSS and AYP
acknowledge support from the Russian Foundation for Basic Research
(grants 08-02-00837 and 09-02-12080) and Rosnauka (Grant NSh
3769.2010.2).
WCGH acknowledges support from the Science and Technology Facilities
Council (STFC) in the United Kingdom through grant number PP/E001025/1.
PSS acknowledges support of the Dynasty Foundation and RF Presidential
Program MK-5857.2010.2.
COH acknowledges support from the Natural Sciences and Engineering
Research Council (NSERC) of Canada.


\label{lastpage}

\end{document}